\documentclass[12pt]{article}
\usepackage{authblk}
\usepackage[utf8]{inputenc}
\usepackage{graphicx}
\usepackage{tikz}
\usepackage{amsmath}
\usepackage{multirow}
\usetikzlibrary{arrows}

\author[1,2]{G. Hern\'andez-Herrera}
\author[1]{A. Navarro}
\author[3,4]{D. Mori\~na\footnote{Corresponding Author: David Mori\~na (dmorina@mat.uab.cat)}}
\affil[1]{Research Group on Psychosocial Risks, Organization of Work and Health (POWAH),		
Unitat de Bioestad\'istica, 
		Facultat de Medicina, Universitat Aut\`onoma de Barcelona}
\affil[2]{Instituto de Investigaciones M\'edicas,
	    Facultad de Medicina,
	    Universidad de Antioquia}
\affil[3]{Barcelona Graduate School of Mathematics (BGSMath), Departament de Matem\`atiques, Universitat Aut\`onoma de Barcelona}
\affil[4]{Department of Econometrics, Statistics and Applied Economics, Riskcenter-IREA, Universitat de Barcelona (UB)}

\title{Regression-based imputation of explanatory discrete missing data}

\date{}

\begin{document}

\maketitle

\begin{abstract}
Imputation of missing values is a strategy for handling non-responses in surveys or data loss in measurement processes, which may be more effective than ignoring the losses. When the variable represents a count, the literature dealing with this issue is scarce. If the variable has an excess of zeros it is necessary to consider models including parameters for handling zero-inflation. Likewise, if problems of over- or under-dispersion are observed, generalisations of the Poisson, such as the Hermite or Conway-Maxwell Poisson distributions are recommended for carrying out imputation. In order to assess the performance of various regression models in the imputation of a discrete variable based on Poisson generalisations, compared to classical counting models, this work presents a comprehensive simulation study considering a variety of scenarios and real data from a lung cancer study. To do so we compared the results of estimations using only complete data, and using imputations based on the Poisson, negative binomial, Hermite, and COMPoisson distributions, and the ZIP and ZINB models for excesses of zeros. The results of this work reveal that the COMPoisson distribution provides in general better results in any dispersion scenario, especially when the amount of missing information is large. When the variable presenting missing values is a count, the most widely used method is to assume that there is equidispersion and that a classical Poisson model is the best alternative to impute the missing counts; however, in real-life research this assumption is not always correct, and it is common to find count variables exhibiting overdispersion or underdispersion, for which the Poisson model is no longer the best to use in imputation. In several of the scenarios considered the performance of the methods analysed differs, something which indicates that it is important to analyse dispersion and the possibile presence of excess zeros before deciding on the imputation method to use. The COMPoisson model performs well as it is flexible regarding the handling of counts with characteristics of over- and under-dispersion, as well as with equidispersion.
\end{abstract}


\section{Introduction}\label{intro}
Missing data are practically unavoidable in research in any field, however their consequences for the validity of research findings are not considered in the majority of cases. Nowadays, many scientific journals emphasise the importance of including information about missing data and the strategy used to handle them (\cite{rezvan2015rise}), and yet it is still not common to find this, and very few publications explicitly describe missing data and the methods used to deal with them. Data can be missing due to non-responses or be caused by problems in study design, or even created deliberately by researchers as part of privacy policies. In order to identify the behaviour of missing data in a sample, in terms of quantity and mechanisms of data loss, the first step is to carry out a statistical description of the observed variables, identify the data lost in each one, and then identify the patterns of data loss which will help to make a decision about a method for handling the missing data. In many cases, researchers resort to using only data which are completely available, ignoring those which are missing when performing analyses; but this decision can generate problems of bias and lack of precision in estimates and affect the power of the study due to the reduction in sample size.

In order to adequately deal with missing data it is necessary to identify the mechanism of data loss (process of non-response) defined as the origin, causes, moment, relationships or characteristics which give rise to the lack of information. Moreover, it is important to establish whether the observations have been lost randomly, or whether their loss is associated with definable causes, and to determine the percentage of missing data in the sample, since depending on these factors, the levels of uncertainty in working with imputed data can vary significantly (\cite{kleinke2011multiple}).

The mechanism of data loss is classified depending on the probability of response: if this is independent of the observed and unobserved data, we say the non-response process is MCAR, missing completely at random. If however it is dependent on the observed data, we say the non-response process is MAR, missing at random. When the process is neither MAR nor MCAR it is termed \textit{informative}, NMAR, not missing at random (\cite{molenberghs1997simple}).

The advantage, statistically speaking, of the mechanism of loss being MCAR is that  conclusions obtained from the analysis of these data can still be valid. The power of the study can be affected by the design, but the parameters estimated are not biased by the absence of data. However, MAR is the most common mechanism and thus it is important to take into account that the probability of data loss is conditionally independent; but if the mechanism is NMAR, this represents a difficulty for imputation, since the estimation of parameters requires knowledge of the model of data loss in order to achieve unbiased estimates.

\cite{little1992regression}, \cite{pigott2001review} and \cite{ibrahim2005missing} have constructed a taxonomy of the most popular methods for handling missing data, which shows that when the data loss mechanism is completely random (MCAR) the methods stochastic regression, multiple imputation, maximum likelihood with Expectation-Maximization (EM) algorithm, Bayesian imputation, and weighted methods produce consistent estimates; when the mechanism of data loss is random (MAR) multiple imputation produces consistent estimates but only under certain conditions, just as with weighted imputation methods, whereas for the non-random mechanism (NMAR) none of the methods produce consistent estimates although it is possible to achieve imputations using some of these methods provided that the correct probability model of data loss can be identified (\cite{lukusa2016semiparametric}).

In many studies, counting variables appear, for example in health research it is common to study episodes of a particular disease and these form the basis of estimates of incidence, when the time-period in which they are observed is also taken into account. Although in the presence of missing data of this type, it would seem logical that the imputation model be based on discrete distributions, such as Poisson or Negative Binomial, in a bio-medical and epidemiological literature review, it was found that very few studies employed these specific methods to impute discrete variables. In practice, the procedures used to handle such missing data were as though the variable had been continuous, or treating it as categorical or ordinal, using polytomic regression techniques, and in other cases using the strategy of applying some normalising transformation to the data, and subsequently using imputation methods for normal data (\cite{landerman1997empirical}).

Much less common is the use of other more sophisticated techniques which tackle specific situations, such as the problem of inflated zeros (\cite{pahel2011multiple}), a situation which may be explained by the lack of software available to carry out imputations of this type. However, some packages have now been developed, for example in R, which allow imputation of these kinds of data (\cite{kleinke2011multiple}) and some studies have stressed the utility of Poisson and Negative Binomial zero-inflated models for epidemiological studies with both cross-sectional and longitudinal designs (\cite{lewsey2004utility}). Few studies use generalised distributions, which are more flexible and take into account problems of over- and under-dispersion, also common in health data: for example no studies are found employing the Hermite distribution, a generalisation of the Poisson distribution, more flexible when there is over-dispersion for the imputation variables of count data. Nor are studies found employing the Conway Maxwell Poisson distribution which permits modeling count data in the presence of over- and under-dispersion.

It is also not common in scientific literature to find an exhaustive examination of missing information in a discrete variable acting as covariate in an analysis. However, this point is crucial in certain contexts such as survival analysis in the presence of recurrent events, where the usual analysis techniques introduced in \cite{Andersen1982} and \cite{PWP1981} require knowledge of the number of previous events suffered by individuals, something which of often unknown (for example events which occurred prior to initiation of a cohort study). The aim of the present study is precisely to assess the performance of methods of imputation of missing data in a discrete covariate, based on generalisations of the Poisson distribution, in comparison to the classical Poisson and Negative Binomial counting methods, in different scenarios of dispersion and nature of response variable and within a framework of multiple imputation, following the recent recommendations in many scenarios like confirmatory clinical trials (\cite{CommitteeforMedicinalProductsforHumanUseCHMP2010}). Although applied researchers were reluctant to using multiple imputation methods until recently, the implementation in most used data analysis software has increased their popularity in the latest years.
\cite{Kim2015} present an alternative based on regression models, but accounting only for continuous covariates. The considered regression models are described in the next section and their performance on real lung cancer clinical trial data and on a comprehensive simulation study are analysed in Section \textit{Results}.

\section{Methods}\label{methods}
In this section we present some discrete variable regression models, on the basis of which to carry out imputation of missing data. The classical counting models including Poisson, negative binomial (parameterised in terms of its mean $\mu$ and dispersion index $d$) and their zero-inflated versions are described in the supplementary material, and only the less known distributions (Hermite and Conway-Maxwell Poisson) are presented here. These distributions are very flexible and may be adapted and used in any scenario of dispersion. A comprehensive description of most common count data modeling strategies with special focus on dispersion issues can be found in \cite{Hilbe2011}. In order to carry out imputation of missing data using the regression models described in this section, two phases are required; firstly, a generalised linear model (GLM) is fitted using the covariate X as response and the response Y as covariate, based on the corresponding distribution (Poisson, NB, ZIP, ZINB, Hermite or COMPoisson). Imputed values are randomly sampled from the corresponding distribution with the parameters obtained in the previous step, including random noise generated from a normal distribution in all cases. In order to produce proper estimation of uncertainty, the described methodology can easily be extended to a multiple imputation framework. The results reported in Section \textit{Results} correspond to the combination of $m=5$ imputed data sets, according to the well known Rubin's rules (\cite{Rubin1987}) and based on the following steps in a Bayesian context:
\begin{enumerate}
 \item Fit the corresponding count data model and find the posterior mean and variance $\hat{\beta}$ and $V(\hat{\beta})$ of model parameters $\beta$.
 \item Draw new parameters $\beta^*$ from $N(\hat{\beta}, V(\hat{\beta}))$.
 \item Compute predicted scores $p$ using the parameters obtained in the previous step (the actual expression depends on the count data model).
 \item Draw imputations from the corresponding count data distribution and scores obtained in the previous step.
\end{enumerate}

The performance of these models in different scenarios will be compared in the next section. The simulation strategy is described in detail in Section \textit{Simulation study}.

\subsection{Hermite distribution}
The Hermite distribution results from the summation of two independent Poisson variables $X_{1}$ and $X_{2}$: $Y= X_{1}+mX_{2}$. It is very useful for modeling count data with a multimodal distribution and with overdispersion (\cite{kemp1965some}).

The probability generating function (PGF) for Hermite distribution is:
\begin{equation}
P(s) = \exp(a_1(s-1)+a_m(s^m-1)).
\end{equation}\label{pgf:gh}

Setting the positive integer $m\ge 2$ (the \textit{order} or \textit{degree} of the distribution), the domain of the parameters is $a_1>0$ and $a_m>0$.

As $a_m$ tends to zero, this distribution tends to a Poisson.

The PGF $P(s)$ corresponds to the PGF of $X_1+mX_2$, where $X_i$ are independent random variables following a Poisson distribution with population mean $a_1$ and $a_m$ respectively (\cite{Puig2003}) The functions for calculating probabilities and fitting Hermite regression models have recently been implemented in R (\cite{Morina2015}).

The Hermite distribution is said to be zero-inflated with respect to the Poisson distribution, because the probability of the variable taking value zero under Hermite is greater than under Poisson, when the two distributions have the same mean. This characteristic of the Hermite distribution allows proposing the use of a Hermite regression model for count variable with an excess of zeros, instead of using a classical Poisson regression model. 

\subsection{Conway-Maxwell Poisson distribution}
A generalisation of the Poisson distribution, taking account of dispersion in the data. The regression model based on the Conway-Maxwell Poisson distribution (COMPoisson) allows modeling count data in three dispersion scenarios: equidispersion, overdispersion and underdispersion, and is therefore an interesting alternative from the point of view of the present study.

The probability distribution function is:
\begin{equation}
P(x, \lambda, \nu)=\frac{\lambda^{x}}{(x!)^{\nu}Z(\lambda,\nu)}
\end{equation}
where $\lambda=E(x^{\nu})$ with $\nu$ the dispersion parameter and
$Z(\lambda,\nu)= \sum_{j=0}^{\infty} \frac{\lambda^{\nu}}{(j!)^{\nu}}$ is a normalising constant. If $\nu=1$ the distribution is equidisperse, and it is overdisperse (underdisperse) if $\nu<1$ ($\nu>1$).

The COMPoisson distribution is an extension of Poisson with two parameters which generalises certain discrete distributions, such as the Poisson distribution when $\nu=1$, the Bernoulli distribution with probability $\frac{\lambda}{1+\lambda}$ when $\nu \to \infty$ and the geometric when $\nu=0$ and $\lambda<1$; it was first suggested in 1962 by Conway and Maxwell in \cite{Conway1962} (see\cite{shmueli2005useful} for a recent review of its applications). This distribution may also be seen as a weighted Poisson distribution with weighting function $ w_{y}= (y!)^{1-\nu}$. In this sense, \cite{Ridout2004} compared the COMPoisson with a weighted Poisson where the weights take the following form:
\begin{equation}
W_{y} = \begin{cases}
e^{-\beta_{1}(\lambda - y)} & \text{if $y\leq \lambda $} \\
e^{-\beta_{2}(y - \lambda)} & \text{if $y> \lambda $}
\end{cases}
\end{equation}

There is underdispersion when $\beta_{1}, \beta_{2}>0$, overdispersion when $\beta_{1}, \beta_{2}<0 $ and equidispersion when $\beta_{1}= \beta_{2}=0$.

This is a flexible distribution which can take account of the excessive or insufficient dispersion often found in count data (\cite{chakraborty2016poisson}). The distribution is appropriate for use in imputation of count data when there are dispersion problems, replacing the Poisson distribution. For example, the COMPoisson distribution has been used in linguistics to model word length, to model count data in marketing, and eCommerce, and to model grocery shop sales, among other uses. The capacity to handle different types and levels of dispersion makes the distribution more useful in applications where the level of dispersion may vary (\cite{sellers2012poisson}).

\subsection{Simulation study}\label{simu}
In order to assess the efficiency of the methods considered in this study, we designed a simulation procedure based on the following algorithm.

\begin{enumerate}
	\item Generate a population of size 1,000,000 with two variables. One variable $Y$, following a binomial, normal or Poisson distribution will be used as the response, depending on the scenario considered (binary, continuous or discrete response variable), and a second will be used as the explanatory variable $X$, consisting of a count of the number of events and based on a Poisson, negative binomial or zero-inflated distribution, depending on the scenario of dispersion and zero-inflation. The covariate $X$ is generated first randomly sampling from the corresponding distribution and then the response $Y$ is generated on the basis of a GLM with different intensities of association with the explanatory variable ($\beta = 0.5$, $\beta = -0.3$ and $\beta = -0.5$) although we only report results corresponding to $\beta = 0.5$ because no relevant differences were observed compared to the other values. These values of $\beta$ were considered to keep the association between the response variable and the coviarate within the usual ranges, with a moderate and reasonable strength.
	\item From the generated population, randomly select 1000 samples of size 2000 and generate missings in the explanatory variable using a MAR or MCAR mechanism, and for percentages of 5\% to 30\%. Regarding missings generated via MAR, the criteria was that 80\% of missings corresponded to values of 0 for the binary variable or values below the mean for discrete or continuous variables and 20\% of the missings corresponded to values of 1 for the binary variable or values above the mean for discrete or continuous variables.
	\item For each sample, fit logistic, linear or Poisson regression models depending on the response variable, as follows:
	\begin{itemize}
		\item \textbf{Listwise deletion (lw)}. Fit the regression model eliminating missing values, in other words, using only the information available.
		\item \textbf{Poisson (pois)}. Fit a Poisson regression model using the count variable as response. Based on the estimated coefficients, impute missing values and subsequently fit the regression model, incorporating the set of imputed values.
		\item \textbf{Negative Binomial (nb)}. Similar  to the above procedure, but impute missing values using a Negative Binomial regression model.
		\item \textbf{Hermite}. In this case, use a Hermite regression model to impute the missing values in the discrete variable.
		\item \textbf{COMPoisson (cmp)}. Use Conway-Maxwell Poisson regression model to perform imputation of the missing data.
		\item \textbf{Zero-inflated Poisson (zpois)}. Use a Poisson regression model with zero-inflation to impute the missing values.
		\item \textbf{Zero-inflated negative binomial (znb)}. Use a Negative Binomial regression model to impute the missing values.
	\end{itemize}
\end{enumerate}

The efficiency of the proposed imputation methods was assessed by comparing relative bias with respect to the population parameter, the average length of confidence intervals for the parameter of interest (AIL) and the percentage of coverage. The same procedure was also carried out using samples of size $n = 200$ but the results showed no differences with those reported, and have been omitted. The following dispersion scenarios were considered:

\begin{itemize}
	\item Equidispersion: the explanatory variable was generated following a Poission distribution with parameter $\lambda = 2$. The procedure described was performed using other values for $\lambda$ but the results did not differ from those reported.
	\item Overdispersion: the explanatory variable was generated following a negative binomial distribution with mean $\mu = 2$ and dispersion index $d = 2$. The procedure described was performed using other values for $\mu$ and $d$ but the results did not differ from those reported.
	\item Underdispersion: the explanatory variable was generated following a Poisson distribution with parameter $\lambda = 2$, and  the underdispersion was generated through an iterative procedure, substituting values for the mean at random until a dispersion of $d = 0.5$ was obtained. The procedure described was performed with other values for $\lambda$ and $d$ but the results did not differ from those reported. In this case, the methods based on the Hermite distribution or on zero-inflated distributions were not considered due to convergence issues.
	\item Excess zeros: the explanatory variable was generated following a Poisson distribution with parameter $\lambda = 2$  and  subsequently a random 10\% of values were replaced by zeros. The procedure described was performed using other values for $\lambda$ and other proportions of zero-inflation but the results did not differ from those reported.
\end{itemize}

The tables including the results for all kinds of response variables and all dispersion scenarios are available as supplementary material.
\section{Results}\label{results}
The performance of the considered imputation methods is compared in this section in a real data example from a randomised clinical trial (RCT) of two treatment regimens for lung cancer, first introduced in \cite{Kalbfleisch1980}. The considered variables are survival time (continuous response) and the number of months from diagnosis to randomisation (discrete explanatory variable), in which a quantity of random missing values were introduced 10\%, 20\%, 30\% and 40\%. Although the main interest in practice in a RCT would be to estimate the treatment effect, we focus here on the effect of the discrete covariate over the continuous response, as it was an observational study like the simulation study presented in Section \textit{Simulation study}.

\subsection{Lung cancer data}\label{lung:cancer}
The explanatory variable number of months from diagnosis to randomisation is clearly overdisperse (the variance is 112.62 while the mean is 8.77). Table~\ref{tab1} shows the performance measures for each of the considered imputation methods (all methods included in a multiple imputation framework with $m=5$ imputed data sets), and it can be seen that COMPoisson performs better than the rest of alternative methods regarding any of the considered measures and in any of the missing data scenarios (10\% to 40\%). Using the full data, the estimate is $\hat{\beta}=-0.69$ $(SD=1.28)$. \\
The 95\% confidence intervals include the reference value of $\hat{\beta}$ in all cases.

\begin{table}
\centering
\caption{Average interval length (AIL) for each imputation method. p-values from $\chi^2$ goodness of fit test comparing the full data and imputed distributions.}
\begin{tabular}{cccc}
\hline
\% of missing data    & Model   & AIL & p-value\\
\hline
\multirow{5}{*}{10\%} & Listwise deletion  & 5.54 & 0.014\\
                      & Poisson            & 5.51 & 0.155\\
                      & Negative binomial  & 5.25 & 0.055\\
                      & Hermite            & 5.38 & 0.154\\
                      & COMPoisson         & 5.25 & 0.053\\
\hline
\multirow{5}{*}{20\%} & Listwise deletion  & 5.78 & 0.002\\
                      & Poisson            & 5.79 & 0.158\\
                      & Negative binomial  & 5.43 & 0.121\\
                      & Hermite            & 5.44 & 0.152\\
                      & COMPoisson         & 5.21 & 0.001\\
\hline
\multirow{5}{*}{30\%} & Listwise deletion  & 8.93 & 0.014\\
                      & Poisson            & 9.62 & 0.647\\
                      & Negative binomial  & 8.84 & 0.373\\
                      & Hermite            & 8.34 & 0.186\\
                      & COMPoisson         & 8.00 & 0.123\\
\hline
\multirow{5}{*}{40\%} & Listwise deletion  & 9.79  & 0.157\\
                      & Poisson            & 10.37 & 0.626\\
                      & Negative binomial  & 8.40  & 0.822\\
                      & Hermite            & 8.59  & 0.156\\
                      & COMPoisson         & 8.37  & 0.306\\
\hline
\end{tabular}\label{tab1}
\end{table}

\subsection{Simulation study results}\label{simu:results}
Below we present the results for estimations of the $\beta$ coefficient and standard error of the regression models fitted, handling missing values in the cases of continuous response variables by the methods: listwise deletion (lw), imputation with Poisson regression (pois), imputation with negative binomial regression (nb), imputation with Hermite regression (herm), imputation using the COMPoisson model (comp), imputation with zero-inflated Poisson model (zpois) and imputation with zero-inflated negative binomial model (znb). For the other types of response variable, results are provided as supplementary material. We also present biases in the estimations and true coverage indices of the confidence intervals for the same case in all scenarios.

\begin{figure}[h!]
\centering
  \includegraphics[height=6cm]{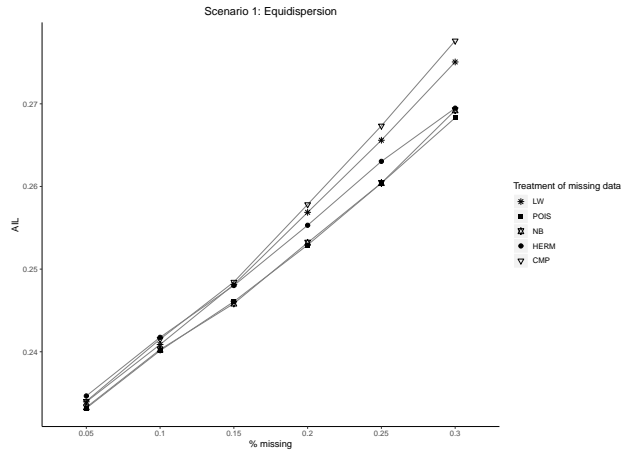}
	\caption{\label{fig:LPIequi}Average length of confidence intervals of the coefficient in the scenario of equidispersion and continuous response variable.}
\end{figure}

\begin{figure}[h!]
	\centering
  \includegraphics[height=6cm]{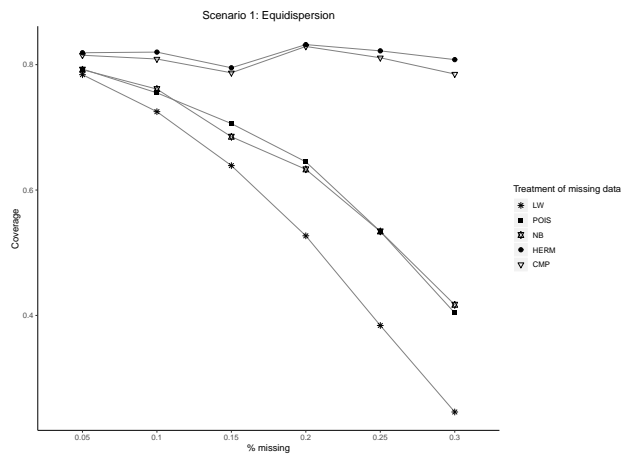}
		\caption{\label{fig:Coverageequi}Coverage of confidence intervals for the coefficient in the scenario of equidispersion and continuous response variable.}
\end{figure}

According to these results, we observe a considerable increase in bias when over 20\% of values are missing, in all scenarios of dispersion and response variable (see Tables S1, S2, S3 and S4) in the Supplementary material. Figure~\ref{fig:BiasequiCont} shows the behaviour of bias for each imputation method under the scenario of equidispersion, and it may be seen that in the estimate using only the information available without imputation (listwise deletion), the bias is greater in comparison to the other methods. The same trends can be seen for the average length of the confidence intervals and their coverage (see Figure~\ref{fig:LPIequi} and Figure~\ref{fig:Coverageequi}).

\begin{figure}[h!]
\centering
  \includegraphics[height=6cm]{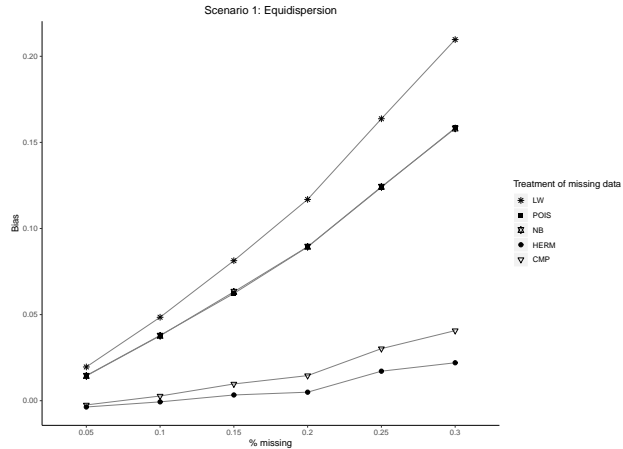}
		\caption{\label{fig:BiasequiCont}Bias in the coefficient estimate for the scenario of equidispersion with continuous response variable.}
\end{figure}

\begin{figure}[h!]
		\centering
		\includegraphics[height=6cm]{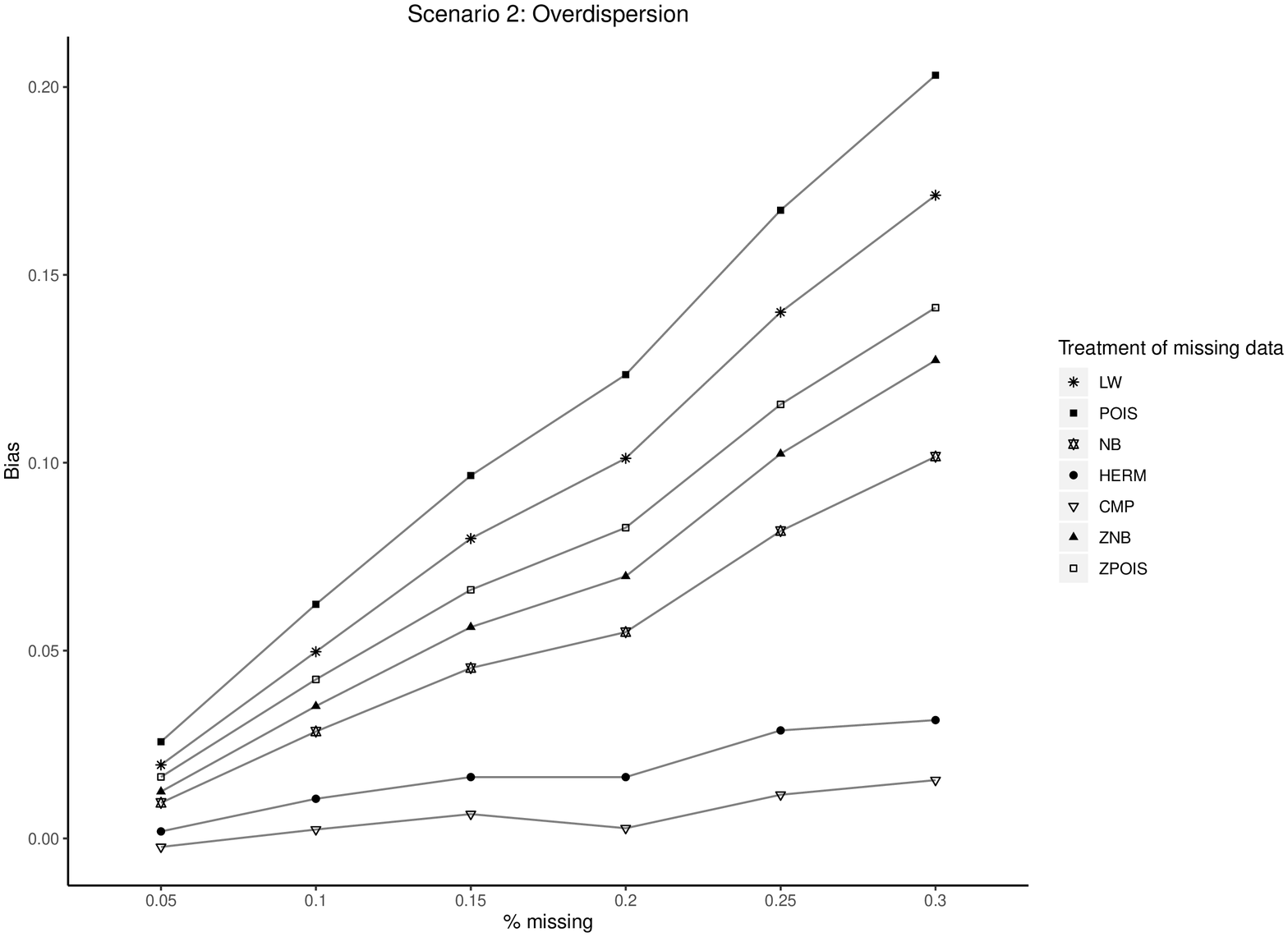}
		\caption{\label{fig:BiasOverd}Bias in coefficient estimation when there is overdispersion and the response variable is continuous.}
	\end{figure}

\begin{figure}[h!]
		\centering
  \includegraphics[height=6cm]{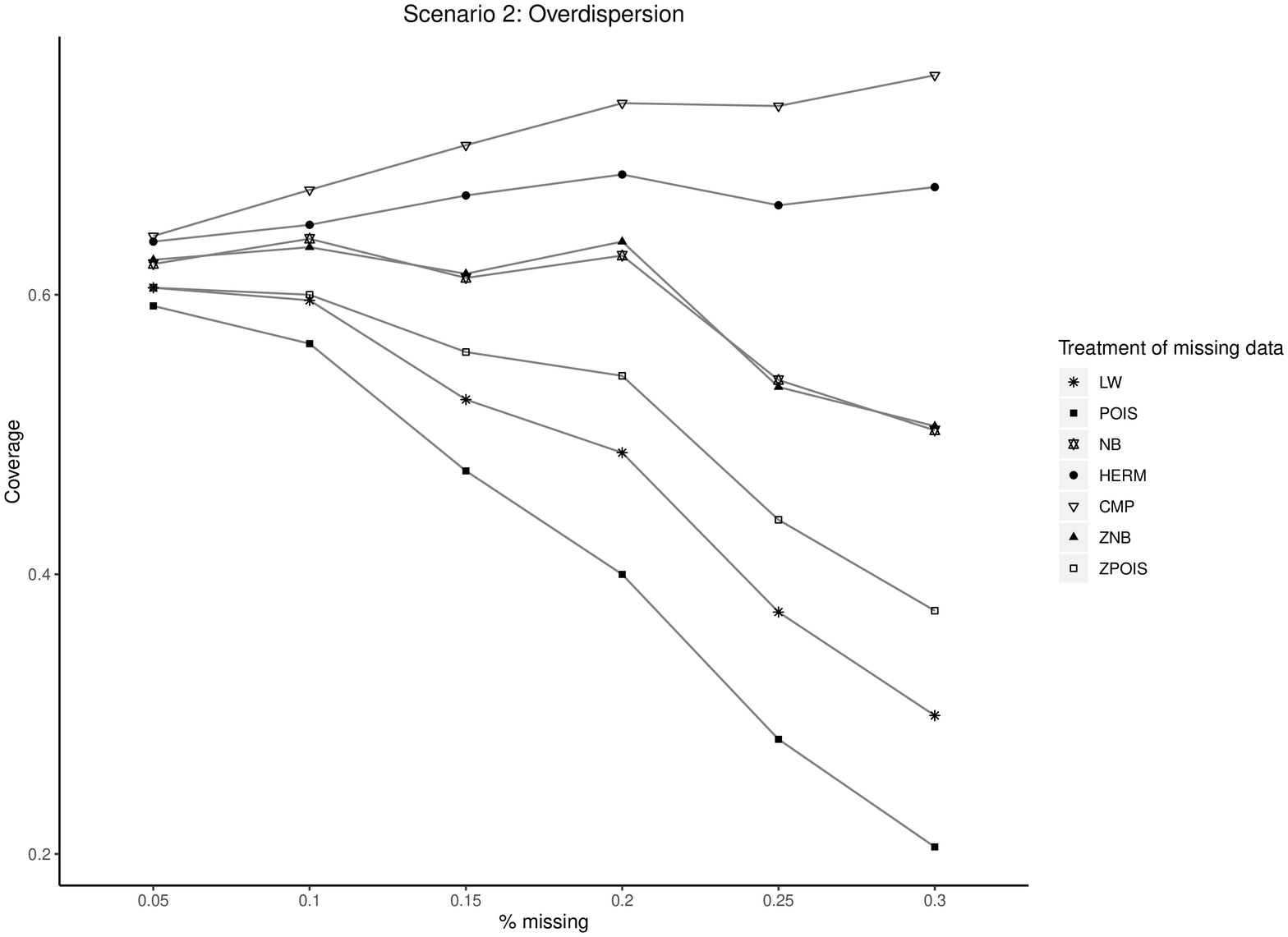}
		\caption{\label{fig:Coverageoverd}Coverage of confidence intervals of the coefficient in the scenario of overdispersion and continuous response variable.}
\end{figure}

In the scenario of overdispersion with a continuous response variable, the behaviour of bias is similar when listwise deletion is used, or when using zero-inflated models to carry out the imputation; lower bias may be seen when imputation is done with the negative binomial or COMPoisson models, as Figure~\ref{fig:BiasOverd} shows, however coverage of the confidence intervals is low compared to the other methods (see Figure~\ref{fig:Coverageoverd}). 

\begin{figure}[h!]
		\centering
\includegraphics[height=6cm]{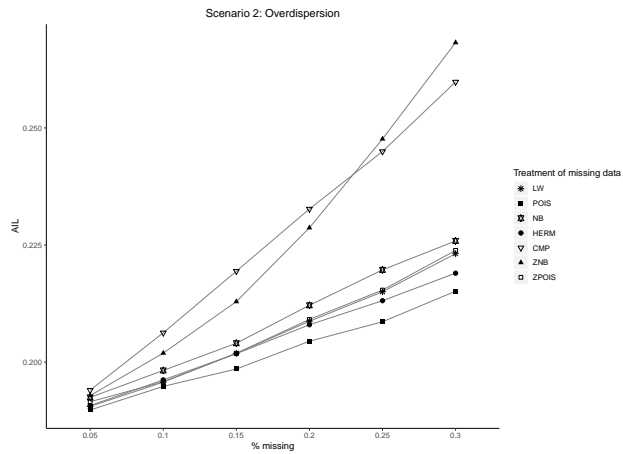}
	\caption{\label{fig:LPIOverd}Average length of confidence intervals of the coefficient when there is overdispersion and the response variable is continuous.}
\end{figure}

\begin{figure}[h!]
	\centering
	\includegraphics[height=6cm]{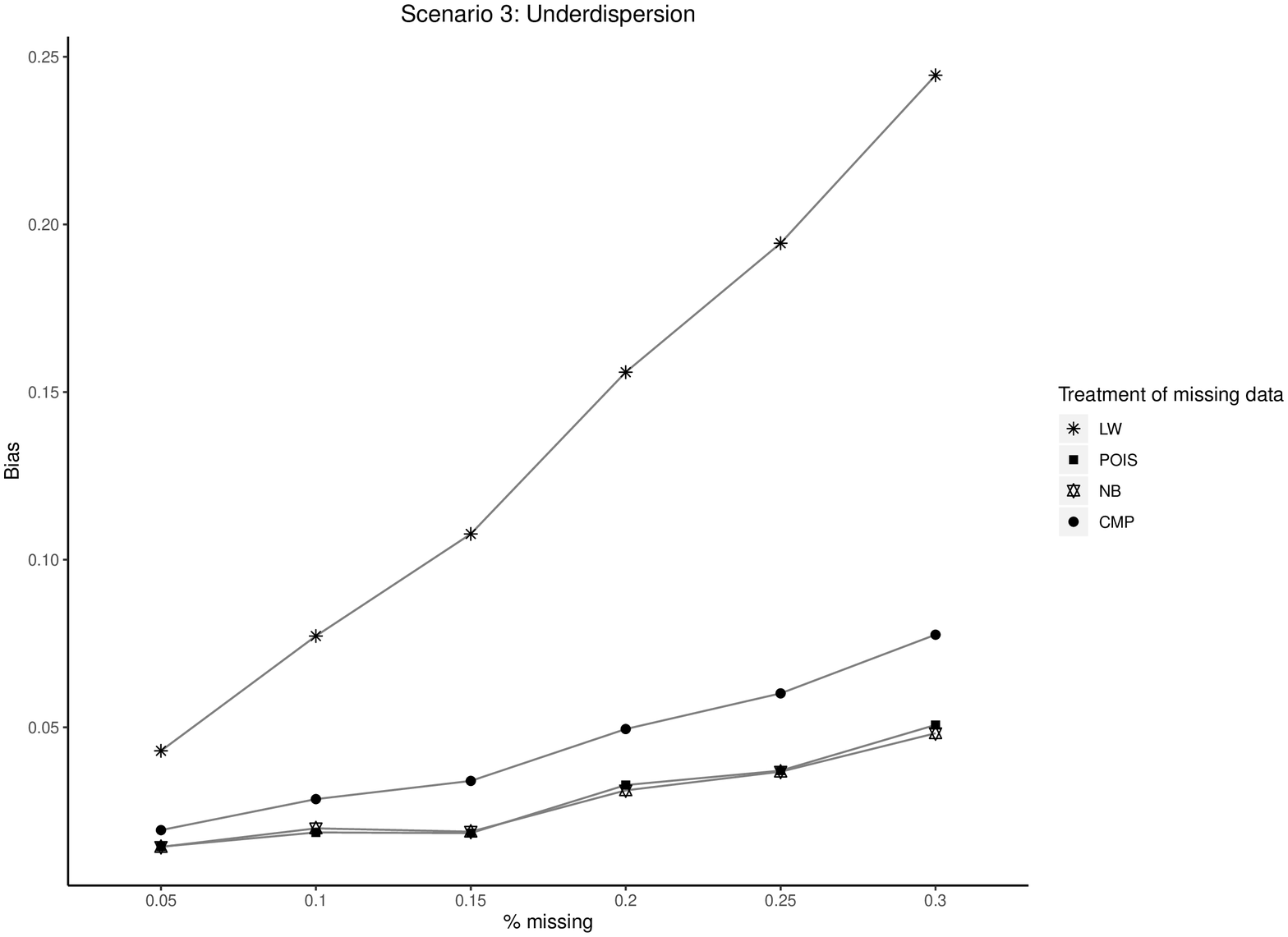}
		\caption{\label{fig:Biasunderd}Bias in estimation of the coefficient when there is underdispersion and the response variable is continuous.}
\end{figure}

\begin{figure}[h!]
	\centering
		\includegraphics[height=6cm]{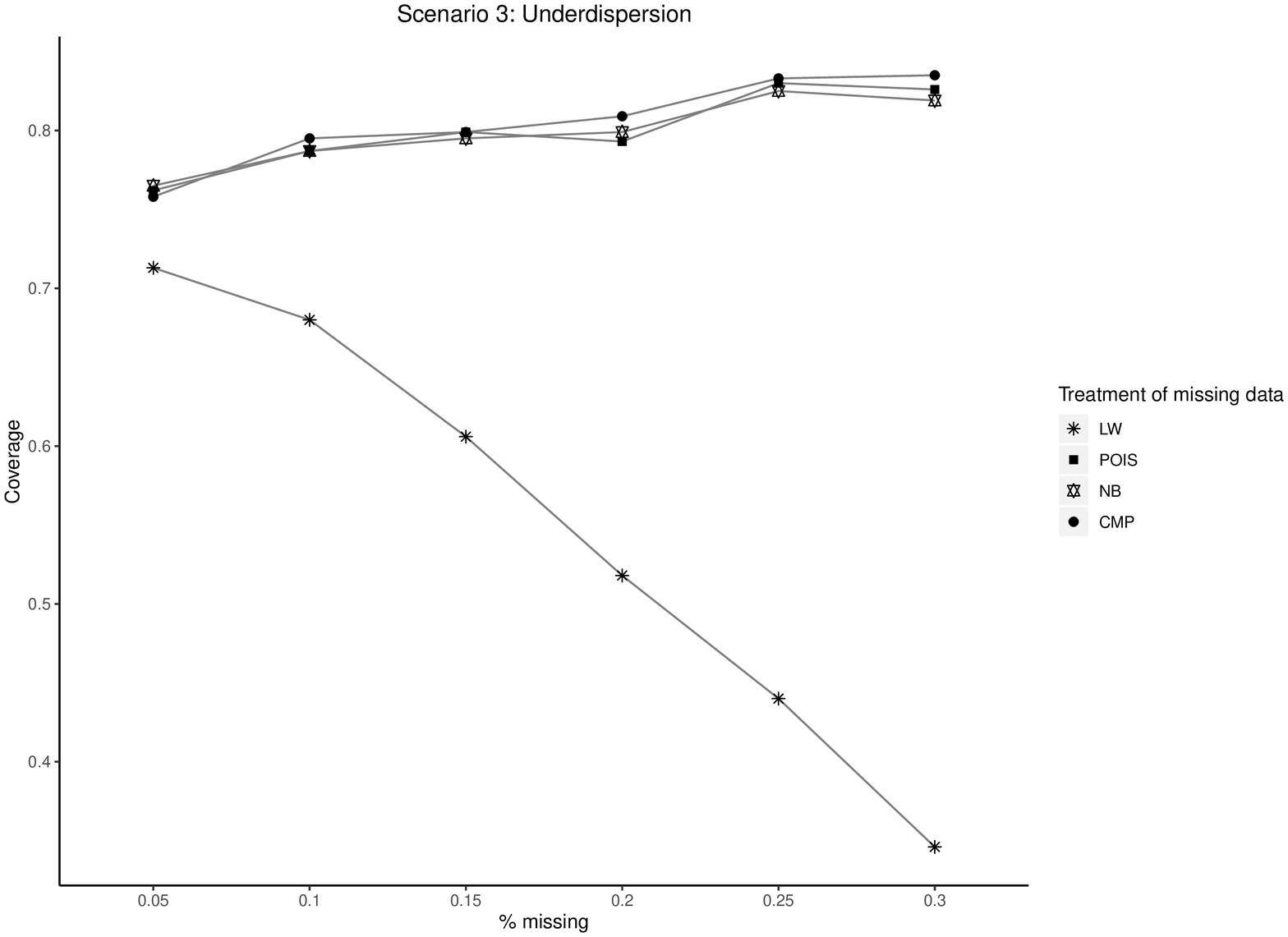}
		\caption{\label{fig:CoberturaInfraCont}Coverage of confidence intervals in the scenario of underdispersion and continuous response variable.}
\end{figure}

In the case of underdispersion the results presented in Figure~\ref{fig:Biasunderd} show that in this scenario the biases are lower when imputing using the Poisson and Negative Binomial regression models (in this scenario it is not possible to obtain the maximum likelihood estimators corresponding to the Hermite distribution) and that the coverage of confidence intervals is greater for these two models (see Figure~\ref{fig:CoberturaInfraCont}).

\begin{figure}[h!]
\centering
\includegraphics[height=6cm]{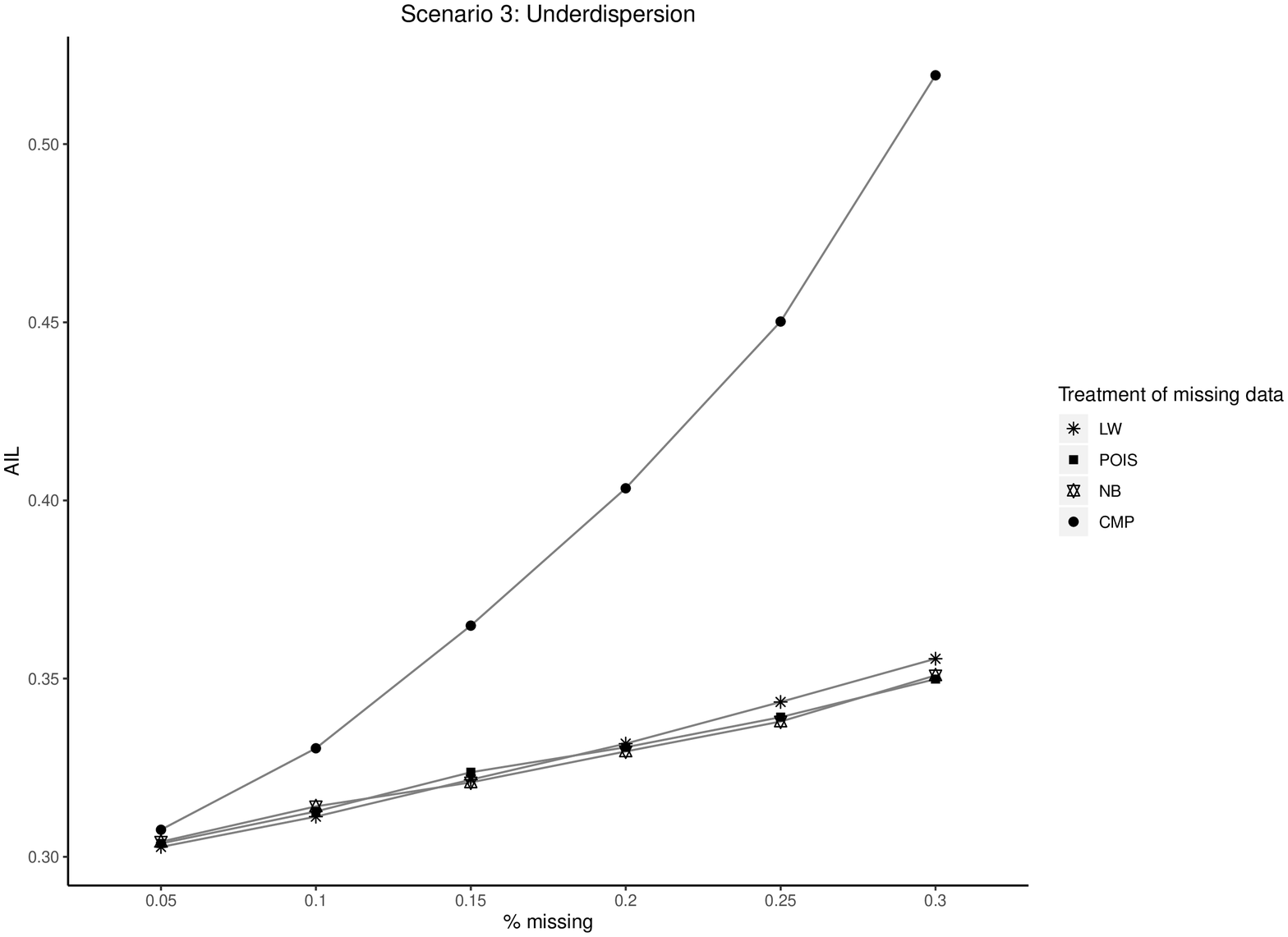}
\caption{\label{fig:LPIInfrad}Average length of confidence intervals in the scenario of underdispersion and continuous response variable.}
\end{figure}

\begin{figure}[h!]
	\centering
 	\includegraphics[height=6cm]{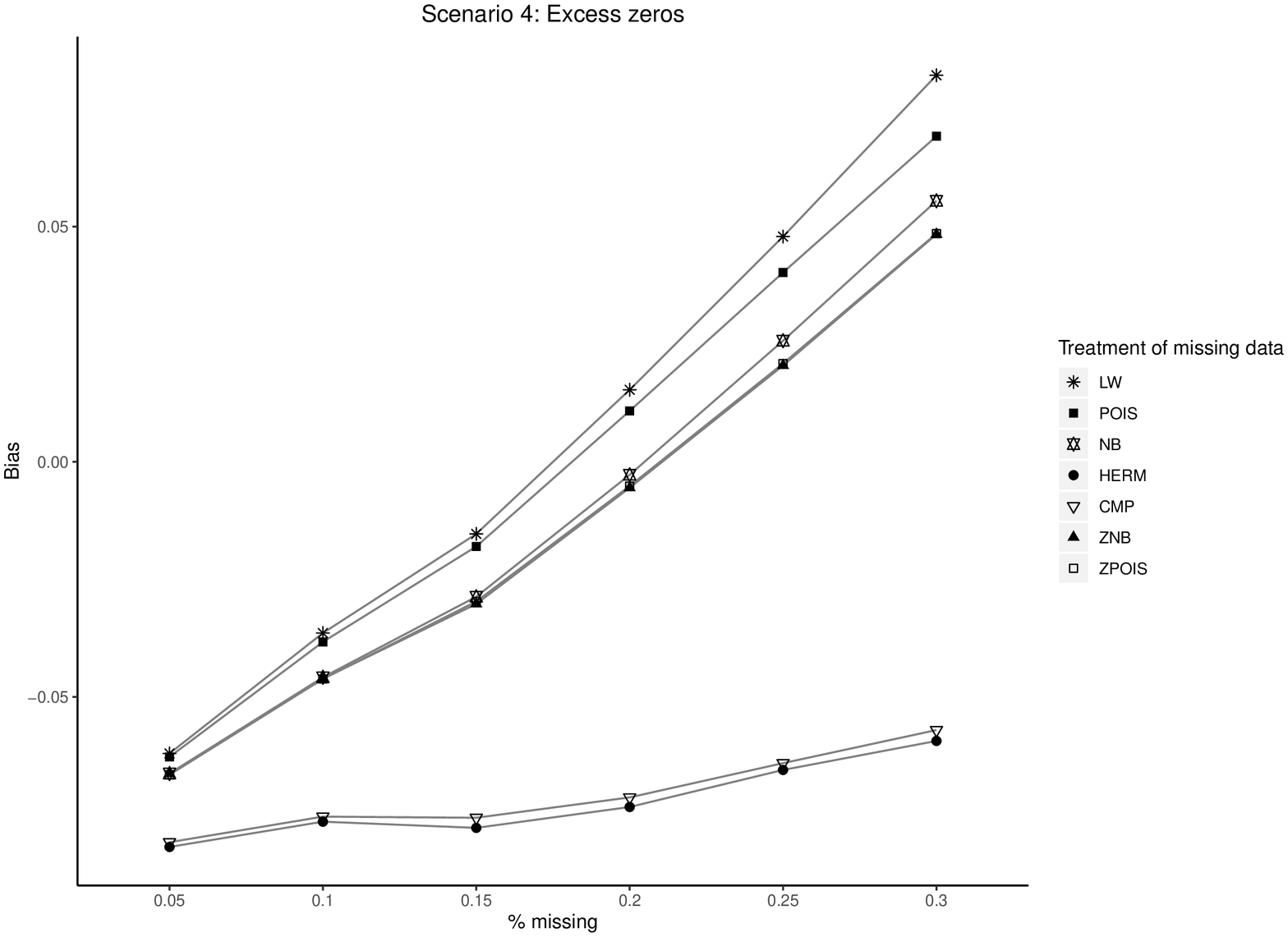}
 	\caption{\label{fig:Biaszeros}Bias in estimation of the coefficient when there is an excess of zeros and the response variable is continuous.}
\end{figure}

When the data present, apart from missing values, an excess of zeros, the results show an improved estimation behaviour using the zero-inflated Poisson and Negative binomial models, as is to be expected, although the performance of the imputation methods considered is in general worse than in the other dispersion scenarios, as may be seen in Figure~\ref{fig:Biaszeros}.

\begin{figure}[h!]
	\centering
	\includegraphics[height=6cm]{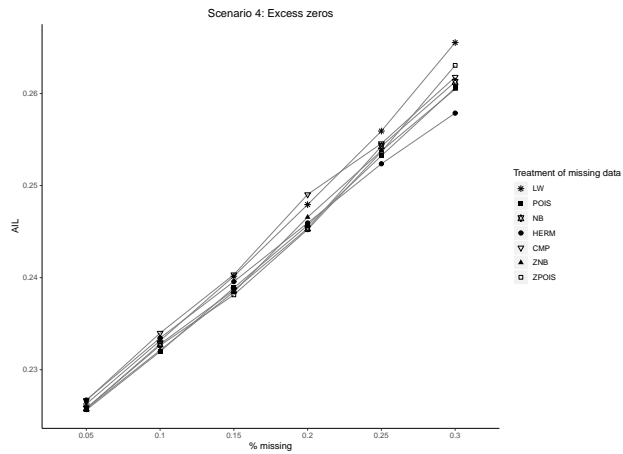}
	\caption{\label{fig:LPIZeros}Average length of confidence intervals  in the scenario of excess zeros and continuous response variable.}
\end{figure}
\begin{figure}[h!]
	\centering
 \includegraphics[height=6cm]{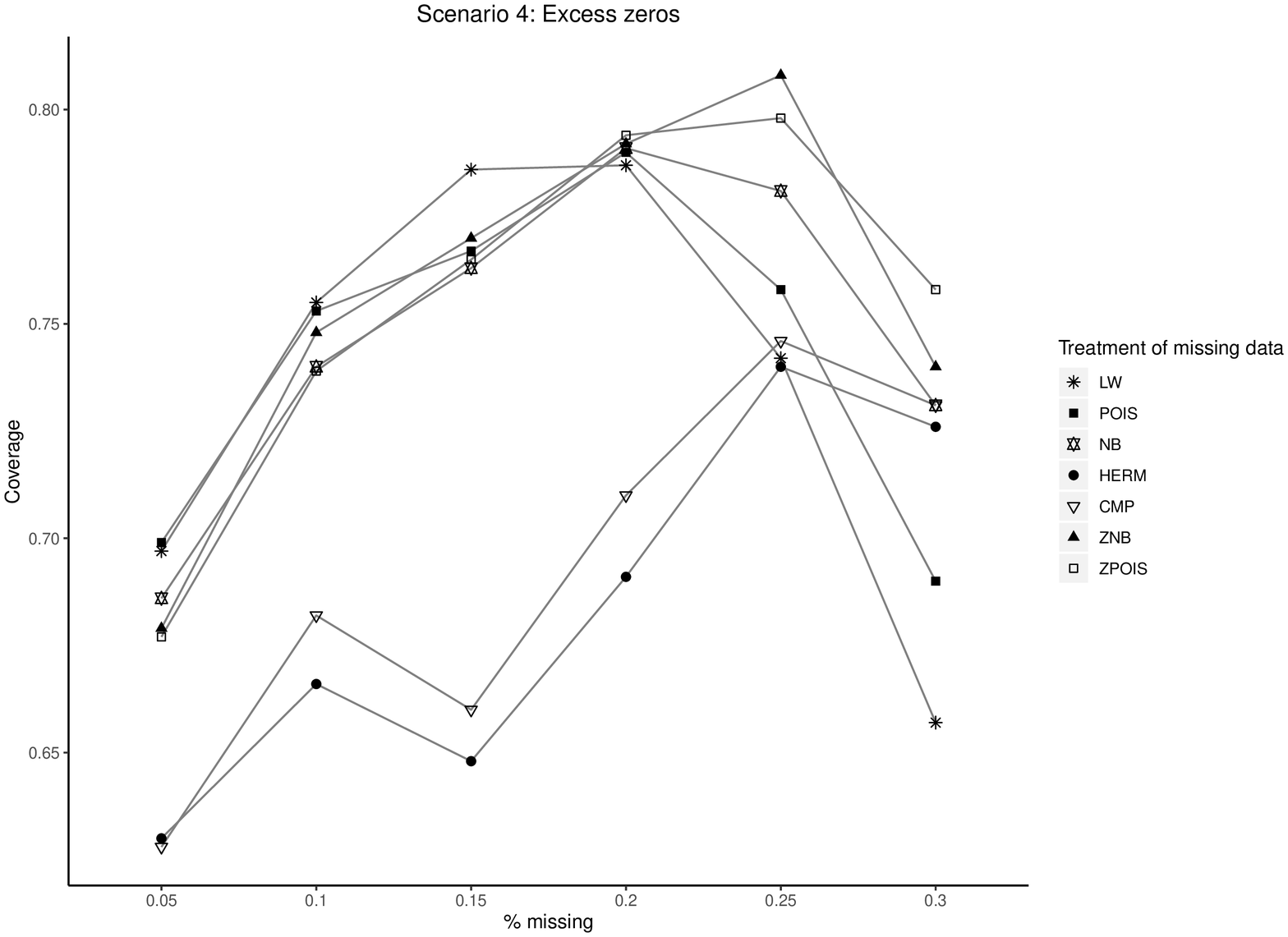}
		\caption{\label{fig:CoberturaZeros}Coverage of confidence intervals when there is an excess of zeros and the response variable is continuous.}
\end{figure}

\section{Discussion}\label{discussion}
In medical research it is common for data to be partially missing and it is necessary to cope with this problem in order to analyse the information in a coherent and consistent manner. Many methods are available which allow this, and proper handling of information which is lacking will depend on choosing the most appropriate one.

When the variable presenting missing values is a count, there are various alternative ways to impute such missing values, which depend on the distribution characteristics of the count variable, particularly the behaviour of the mean and variance. The most widely used method is to assume that there is equidispersion and that a classical Poisson model is the best alternative to impute the missing counts; however, in real-life research this assumption is not always correct, and it is common to find count variables exhibiting overdispersion or underdispersion, for which the Poisson model is no longer the best to use in imputation. If there is overdispersion the Poisson model underestimates the amount of dispersion. In recent years much work has been done on implementation of other counting models, which may be generalisations of the Poisson model and which take over- and under-dispersion into account, as well as the problem of excess zeros (\cite{raghunathan2001multivariate}).

The lung cancer example shows that the COMPoisson distribution provides with a powerful alternative to missing data imputation in a realistic situation of overdisperse discrete explanatory variables, even with a large proportion of missing information, within the framework of multiple imputation, which is now implemented in many standard software and is therefore available to researchers in a straightforward way.

Moreover, in the simulation study we generated different scenarios with a variety of percentages of missing data and three distributions of the response variable: continuous, binary and discrete, and subsequently fitted the respective models with the imputed independent variable using one of the models mentioned. The results of this simulation allow us to affirm that in order to impute a count variable, it is not sufficient to assume the distribution is Poisson; it is necessary to identify the relationship between the variance and the mean of the data, as well as whether the presence of excess zeros might make it appropriate to use specific models for handling missing data in this scenario. Additionally, it can be seen that especially for continuous and discrete response variable when the proportion of missing information is very high (over a 20\%), imputing the missing values can lead to inaccurate results regarding relative bias and extremely low coverage rates.

One of the most unexpected findings is that fitting models using only the available data in some cases produces estimator with less bias than performing imputation of the missing values (\cite{Hernandez2017}), in the case of a dichotomic response variable. Although it is recognised that fitting a model with only the available data may affect the power of the study and produce imprecise estimations, it is evident that the effects on power and precision may be significant if the percentage of missings is low and the mechanism of data loss is completely random.

\section{Conclusions}\label{conclusions}
In several of the scenarios considered the performance of the methods analysed differs, something which indicates that it is important to analyse dispersion and the possibile presence of excess zeros before deciding on the imputation method to use. Specifically, in the scenario of equidispersion and binary response variable, when a logistic regression model is fitted imputing missing values with the MAR mechanism using the different models mentioned we observe, as expected, that the Poisson and Negative Binomial models produce estimations with low bias and acceptable coverage. Moreover, the COMPoisson model performs well as it is flexible regarding the handling of counts with characteristics of over- and under-dispersion, as well as with equidispersion (\cite{sellers2012poisson}). If, however, the count variable is overdisperse, as is often the case in health research, there are various alternatives for performing imputation, the Negative Binomial model being the most recommended. In our results, just as in the case of equidispersion, when the response variable is binary, estimating using available data without performing imputation produces good estimators and with low bias, however it is important to observe that in this case the imputation methods which perform best are those employing zero-inflated models, and the COMPoisson model works very well.
For the case where the variable presents underdispersion we observe that imputation based on Poisson and Negative Binomial regression models perform similarly, although the Negative Binomial can present certain difficulties with convergence in this scenario. Furthermore, regard in size of confidence intervals, it is curious that in fitting these models, the higher the percentage of missings the smaller the confidence intervals (apparently an artifact) whereas the COMPoisson is able to maintain their size, and it is worth emphasising that with listwise deletion size increases, as in the other scenarios. When in addition to having missings the data has excess zeros the estimates of the $\beta$ parameter in the case of binary response are more precise, i.e. they have less bias and greater coverage but, as in the previous scenarios, this result is similar to when no imputation is performed and only the available data is used. For the case of continuous response variable (Table S4) estimates of the $\beta$ coefficient are always smaller than the value of the parameter and coverage of the confidence intervals is acceptable, coverage being even greater for listwise, although at the expense of some confidence intervals being considerably wider.

According to the results of the simulation obtained in this study, the choice of the best method of imputation for count variables depends on various factors such as the amount of missing data, the behaviour of the expected value in relation to the variance, i.e. whether there is equi-, under-, or over-dispersion and the distribution of the response variable. In particular, if data present an excess of zeros, this represents an additional factor to be taken into account when choosing the missing data imputation method. Although in practice the exact distributional form of the incomplete covariates is unknown because, precisely, of the missing information, the behaviour of many phenomena in the context of public health are well established. For instance, it is well known that the risk of suffering a new sickness leave increases with the number of previous events (see \cite{Reis2011} for instance), which would lead to overdispersed data, and the same behavior can be observed with the number of falls suffered by long-term centers residents (as in \cite{Navarro2009}).

\section*{Acknowledgements}
  David Mori\~na acknowledges financial support from the Spanish Ministry of Economy and Competitiveness, through the Mar\'ia de Maeztu Programme for Units of Excellence in R\&D (MDM-2014-0445) and from Fundaci\'on Santander Universidades.

\bibliographystyle{plain}
\bibliography{refs}

\end{document}